\documentclass[a4paper,fleqn]{cas-dc}
\usepackage[square,numbers]{natbib} 
\usepackage{setspace}
\usepackage{graphicx}
\usepackage{amsmath}
\usepackage{amssymb}
\usepackage{bm}
\usepackage{subcaption}
\usepackage{microtype}
\usepackage[ruled,vlined ]{algorithm2e}
\begin{document}
\shorttitle{Lung sound classification using ML}
\shortauthors{Elsetrønning et~al.}
\title [mode = title]{On the effectiveness of signal decomposition, feature extraction and selection on lung sound classification}                      
\author[1]{Andrine Elsetrønning}
\ead{andrine.els@stud.ntnu.no}
\address[1]{Norweigan University of Science and Technology, Elektro D/B2, 235, Gløshaugen, O. S. Bragstads plass 2, Trondheim, Norway}
\author[1,2]{Adil Rasheed}
\ead{adil.rasheed@ntnu.no}
\cormark[1]
\address[2]{Mathematics and Cybernetics, SINTEF Digital, Klæbuveien 153, Trondheim, Norway}
\ead[url]{www.adilrasheed.com}
\author[3]{Jon Bekker}
\ead{jon@dedex.ai}
\address[3]{dedeX, Øverbergveien 12B
1397, Nesøya, Norway}
\author[4]{Omer San}
\ead{osan@okstate.edu}
\ead[url]{www.cfdlab.org}
\address[4]{Oklahoma State University, School of Mechanical and Aerospace Engineering, 201 GAB, Stillwater, Oklahoma 74078 USA}

\begin{abstract}
\paragraph{}
    Lung sounds refer to the sound generated by air moving through the respiratory system. These sounds, as most biomedical signals, are non-linear and non-stationary. A vital part of using the lung sound for disease detection is discrimination between normal lung sound and abnormal lung sound. In this paper, several approaches for classifying between no-crackle and crackle lung sounds are explored. Decomposition methods such as Empirical Mode Decomposition, Ensemble Empirical Mode Decomposition, and Discrete Wavelet Transform are used along with several feature extraction techniques like Principal Component Analysis and Autoencoder, to explore how various classifiers perform for the given task. An open-source dataset downloaded from Kaggle, containing chest auscultation of varying quality is used to determine the results of using the different decomposition and feature extraction combinations. It is found that when higher-order statistical and spectral features along with the Mel-frequency cepstral coefficients are fed to the classier we get the best performance with the $k$-NN classifier giving the best accuracy. Furthermore, it is also demonstrated that using a combination of feature selection methods one can significantly reduce the number of input features without adversely affecting the accuracy of the classifiers.
\end{abstract}

\maketitle

\section{Introduction}
\label{section:introduction}
    Data-driven machine learning (ML) algorithms are revolutionizing the world. They have emerged as one of the most important enablers for Digital Twins (DT) \cite{rasheed2020digital}. Although DT of a human body is a far-fetched idea at the moment, there is no denying the fact that cheap sensors, storage and communication technologies will facilitate not only on demand data collection but also sharing in real time. Recently ML has achieved appreciable success in the field of medicine like in detection of Glaucoma \cite{ran2020dli}, analysis of electrocardiogram \cite{MINCHOLE2019S61}, detection of diabetes mellitus and self-management \cite{CHAKI2020} to name a few. More related to the topic of investigation in this work, the authors in \cite{CHEN2021112799} have presented a system equipped with ML algorithm, to continuously monitor respiratory behaviors. The respiratory behaviours are characterized by lung sound (LS) which refers to the specific sound produced by air moving through the respiratory system. To record these sounds, currently, a stethoscope is usually placed across the chest/thorax \cite{MERUVIAPASTOR201615}. Using chest auscultation one can gain knowledge about the state of the lungs in a quick and inexpensive manner. If one can accurately detect abnormalities in the lungs, just by processing the audio, this could be a helpful tool for medical practitioners since they use LS as one of many sources of information about a patient which in many cases guide further investigation or management. 

    Several approaches for discriminating between normal LS and abnormal LS have been presented throughout the years. To identify the knowledge gap, a collection of some of the most interesting findings is presented. In \cite{kandaswamy2004neural} a classification method was built upon firstly denoising the signal using wavelet shrinking, then normalizing the LS, followed by decomposition by Discrete Wavelet Transform (DWT), statistical feature extraction from the wavelet coefficients, and classification using an Artificial Neural Network (ANN). With 6 different labels; normal, wheeze, crackle, squawk, stridor, or rhonchus, an accuracy of 94.02\% was reported. Rational dilation wavelet transform is proposed as a decomposition method in \cite{ulukaya2017overcomplete}. The Shannon entropy, standard deviation, energy, maximum/minimum, mean, and skewness/kurtosis were computed on each frequency sub-band. Furthermore different classifiers were tested and compared, with a Support Vector Machine (SVM) showing the most promising results, with an accuracy of 95.17\%. When using ensemble learning the accuracy could be amplified to 97.67 \%. The work presented in \cite{palaniappan2013machine} gives a systematic overview of the recent trends in LS classification, by means of ML. Clearly, the trend, giving the most accurate results, is to use either ANN or k-NN to classify the LS. The paper points out wavelet features as a common features in computer-based LS analysis. In \cite{DBLP:journals/corr/GronnesbySHMB17} the authors focus on extracting a 5-dimensional feature vector from small windows in the audio files. Four of the features were taken from the time domain, while one feature was from the spectrum domain. A precision of 86 \% and a recall of 84 \% were obtained for classifying crackles. 

    As the dataset to be utilized in this report has the advantage of being open-source, made for a scientific challenge, there already exist several papers about approaches researchers have investigated.  The challenge was to classify between crackle, wheeze, both, or none. A training dataset was made public, while the testing-set was held private for later comparison of the results. A combination of Mel-Frequency Cepstral Coefficient (MFCC) features with a Hidden Markov Model (HMM) was presented in \cite{10.1007/978-981-10-7419-6_7}, here a sensitivity of 0.423 and a specificity of 0.566 was attained. In \cite{10.1007/978-981-10-7419-6_8} a non-linear spectral feature extraction method was utilized. Here a non-linear wavelet-based decomposition was applied. The features were extracted from the decomposed bands, by means of the Short Time Fourier Transform. The extracted features along with an SVM for classification gave a selectivity of 0.489 and a specificity of 0.778. The research in \cite{chambres2018automatic} focused on demonstrating the importance of adopting a patient-level point of view for classification. The classification part focused on deciding if adventitious sounds are detected or not, this is accomplished by using a boosted decision tree. 

    There is a noticeable trend in the papers presented, of firstly applying different preprocessing steps, that includes noise reduction, down-sampling, and slicing. Secondly, several papers decompose the LS into sub-bands, which could carry more relevant information about the abnormalities. Conclusively, for classification, a collection of classifiers are tried and tested. The combinations of the distinctive steps performed in the literature are endless, yet little research has been conducted where the various combinations are investigated extensively. No paper has been found, giving a comprehensive study of decomposition, feature extraction, feature selection and classifier combinations, thus the focus of this work is placed here. The main objective of this work is to find an efficient and robust approach for classifying a breathing cycle as either no-crackle or crackle. In order to do so, we evaluate the combined performance of various time-series decomposition techniques, feature extraction and selection techniques with the state-of-the-art machine learning classification methods. 
  
    The paper starts with Section \ref{section:theory} giving a quick overview of the theory behind LS, time-series decomposition methods, feature extraction / selection methods, and supervised / unsupervised machine learning methods used in this work. Details regarding the data and its pre-processing are given next in Section \ref{section:data}. Details regarding the set up for the analysis is given in Section \ref{section:setup}. Results are finally presented in Section \ref{section:results} followed by conclusion and future research directions in Section \ref{section:conclusions}.

\section{Theory}
\label{section:theory}
\begin{figure}
    \centering
    \includegraphics[width = \linewidth]{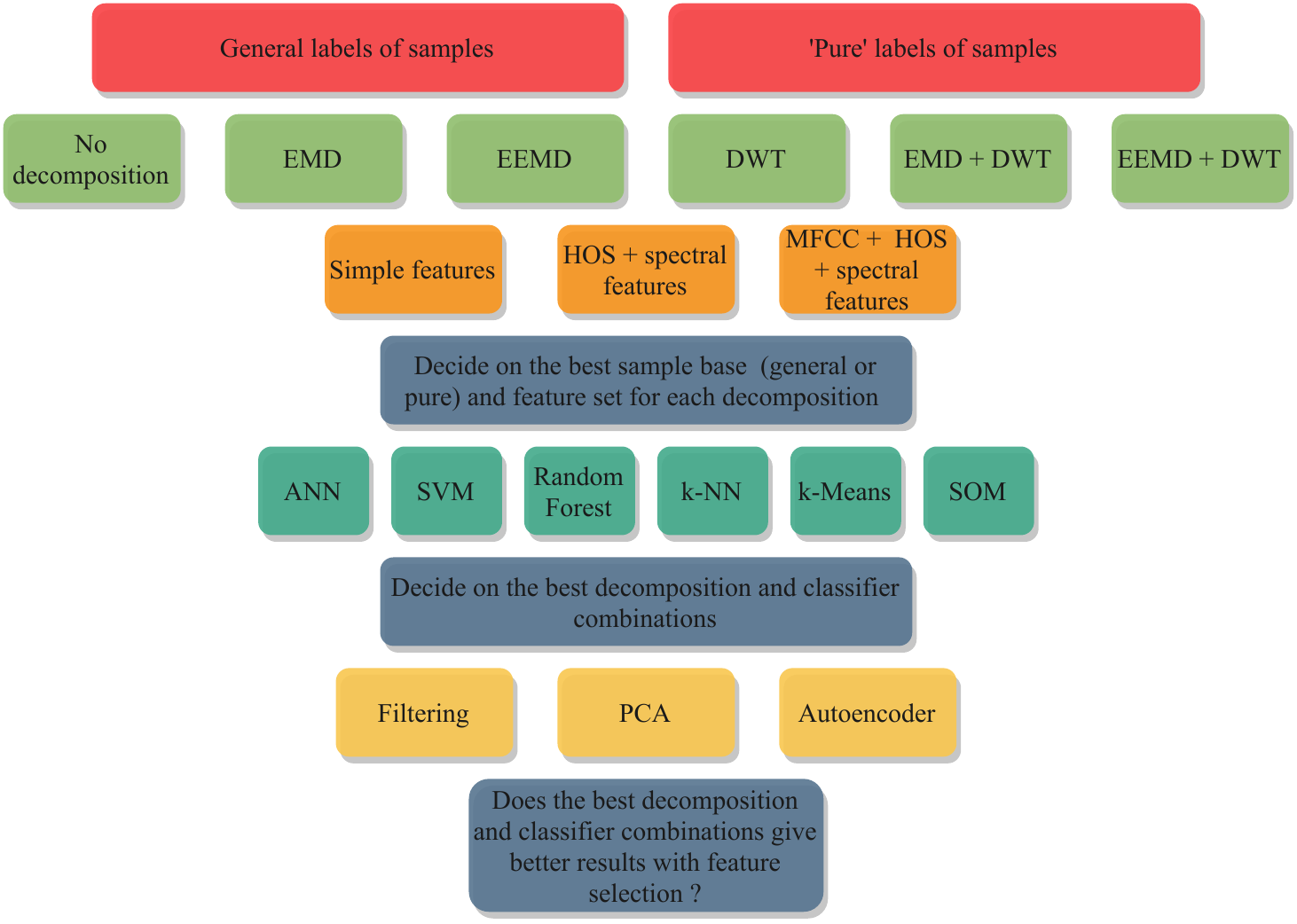}
    \caption{Work flow}
    \label{fig:workflow}
\end{figure}
In this section, we give an overview of the knowledge / theory that forms the basis of this work. Figure \ref{fig:workflow} gives an overview of the different steps involved in the analysis. Since it is difficult to apply machine learning algorithms directly on the audio data, the signal was firstly decomposed, before extracting relevant features. The relevant algorithms are briefly described in the following sections.  

\subsection{Lung Sound and Disorders}
One breathing cycle is defined as the sequence of events in which a human/animal inhales and exhales. For humans, the average respiratory cycle lasts for about 5 seconds \cite{RespCycle}. \textit{Crackles} are discontinuous, explosive, and non-musical adventitious respiratory sounds that occur frequently \cite{KaggleData}. The appearance of crackles is often an early sign of respiratory disease. It is difficult to detect crackles in the power-spectrum since the frequencies range from 200-2000Hz, and the crackles take on a transient waveform \cite{SENGUPTA2016118}. \textit{Wheezes} are music-like adventitious sounds that usually last more than 100 ms, with a frequency greater than 100 Hz \cite{7443768}. Wheezes usually indicate some kind of airway obstruction.  

\subsection{Decomposition Methods}
The audio signals are first decomposed using the following two algorithms. 

\textit{Empirical Mode Decomposition (EMD):} Is a method for decomposing a time-series that, unlike Fast Fourier Transform (FFT), does not assume periodicity in the signal. Being an empirical algorithm, EMD can decompose non-stationary and non-linear signals, hence the method is ideal for analyzing biomedical signals \cite{zeiler2010empirical}. EMD breaks down signals into Intrinsic Mode Functions (IMF’s). After conducting EMD analysis, the original signal is represented as a sum of its IMF's as well as the final residual function as in
equation (\ref{eq: EMD IMF res}).
    \begin{equation}\label{eq: EMD IMF res}
        x[n] = \sum_{l=0}^{L} IMF_l[n] + r[n]
    \end{equation}
    
In Equation (\ref{eq: EMD IMF res}) $x[n]$ is the original input signal, $IMF_l$ is the $l$-th level extraction of an IMF and $r[n]$ is the residue. EMD is heuristic, hence it is subject to some short-comings  \cite{zeiler2010empirical}. One common problem with EMD is mode mixing, meaning that the same information can leak into many IMF's \cite{damavsevivcius2017imf}. When mode-mixing occurs, the EMD algorithm is rendered unstable, since any small perturbation in the data can lead to a totally different set of IMF's. In Ensemble Empirical Mode Decomposition, EEMD, mode-mixing is reduced, leading to improved separation of frequencies. EEMD exploits the properties of white noise, to ensemble an average of IMF's, according to Equation \ref{eq:EEMD}.
\begin{equation}\label{eq:EEMD}
    IMF_l[n] = \frac{1}{W}\sum_{w=1}^{W} IMF_{lw}[n] + r[n]
\end{equation}

Above, $W$ is the number of white noise iterations, and $IMF_{lw}$ is the IMF at the $w$-th iteration of adding white noise. More details can be found in \cite{zeiler2010empirical}.
    
\textit{Discrete Wavelet Transform:} The wavelet transform (WT) can be viewed as the projection of a signal into a set of basis functions, which are called wavelets \cite{KEHTARNAVAZ2008175}. These wavelet functions are dilated, translated, and scaled versions of a common mother wavelet, together forming an orthogonal set of basis functions. WT is used as a method to obtain knowledge of the time-frequency localization. Discrete Wavelet Transform (DWT) is a method for denoising or decomposing a signal, using WT. In DWT the signal is level-by-level passed through a half-band low-pass filter, giving the approximation coefficients, and a half-band high-pass filter to get the detail coefficients \cite{157221}. After filtering, the signal is down-sampled by two, following the Nyquist sampling criteria. 
 

\subsection{Feature Extraction} \label{sec: Theory Feature Extraction}
The section below contains an overview of the theory behind some of the most common methods for feature extraction. Features are extracted from the decomposed signal. 

\textit{Simple statistical features:} With inspiration from \cite{kandaswamy2004neural}, a simple statistical feature selection is composed of the mean ($mu$), standard deviation ($\sigma$), as well as the minimum and maximum of the decomposed time-series vectors. 

\textit{Higher order statistical features:} In \cite{ulukaya2017overcomplete}, a feature set composed of basic statistical features such as mean, standard deviation, minimum, maximum, entropy and energy were presented, as well as higher order statistics, such as skewness, kurtosis, Shannon entropy and Energy. Since these features showed such promising results, they will be employed as a more complex feature extraction. 

\begin{align}
    S &= - \sum_{n = 1} d_n^2 log(d_n^2)  \label{eq: Shannon Entropy def}
    \\[1em]
    E &= \frac{1}{N} \sum_{n = 1}^{n = N} |d_n|^2 \label{eq: Energy def}
\end{align}

In Equations \eqref{eq: Shannon Entropy def}-\eqref{eq: Energy def} , $S$ is the Shannon Entropy, $E$ is the energy, $d$ is the decomposed signal vector and $N$ is the number of samples in one data-vector $d$.

\textit{Spectral features:} The spectral representation of a signal expresses the distribution of energy over a set of frequencies.  In \cite{sharma2020trends} a common trend of extracting spectral features from audio signals is discussed. Spectral features are often extracted from the Short-Time Fourier Transform (STFT) of the signal, where the spectrogram is divided into frames of a set length. For the proposed approach the spectral features are extracted from the STFT of the decomposed signal bands $d$. The spectral features are extracted from $d$ to explore the energy distributed in the decomposed bands. The following features are computed from every frame: spectral centroid,  spectral bandwidth, spectral contrast, spectral roll off, and zero crossing rate. 

\begin{align} 
    C [\tau] &= \frac{\sum_k STFT[k,\tau]  f_k }{\sum_j S[j,\tau]} \label{eq: spec centroid}
    \\[1em]
    B [\tau] &= (\sum_k STFT[k,\tau]  (f_{k\tau} - C[\tau]) ^p )^{1/p} \label{eq: spec bandwidth} 
\end{align}

In equations (\ref{eq: spec centroid}) - (\ref{eq: spec bandwidth}), $C$ is the spectral centroid, $B$ is the spectral bandwidth, $\tau$ is a time frame ,$f_k$ is the frequencies of the rows in $STFT$, and  $p$ is the order of the spectral bandwidth. 

\textit{Spectral contrast} is computed by dividing each frame of the spectrogram into sub-bands, and for each sub-band compare the mean energy in the top quantile, to the bottom quantile. \\
\textit{Spectral roll-off} is defined as the center frequency for a spectrogram bin, where at least a given percent of the energy of the frame is contained in this bin and below. \\
\textit{zero-crossing-rate} is straight forward the number of zero crossings divided by the fame length. 

For an in-depth explanation of how the spectral features are computed, see \cite{mcfee2015librosa}. 

\textit{Mel-Frequency Cepstrum Coefficients (MFCC) features}, refer to mathematical coefficients for sound modelling. It is found that the human perception of frequency components in sounds from speech signals does not follow a linear scale \cite{tiwari2010mfcc}. In order to find a more fitting scale, the Mel scale was created. The Mel scale maps frequencies from the Hertz scale, into a scale where listeners judge the pitches to be in equal distance from one another. This mapping comes from the observation that the perceptual distance between lower frequencies is greater than the perceptual distance between higher frequencies. Equation (\ref{eq: mel-scale}) gives the mapping from frequency to Mel scale. 
\begin{equation}\label{eq: mel-scale}
    M(f) = 2595 log_{10} ( 1 + f/700)
\end{equation}
where $M$ is the Mel-frequency and $f$ is the frequency in Hertz. For a in depth explanation of how MFCC-features can be extracted from a audio time-series, see \cite{tiwari2010mfcc}. 

\subsection{Feature Selection}
\label{subsec:featureselection}
Generally including too many irrelevant features during training can make a machine learning model overly complex and hence prone to overfitting. We use three feature selection techniques to either directly select the most important features or to engineer new features and get rid of redundancies.

\textit{Filtering Approach:} This approach results in discarding certain features based on certain statistical criteria like removing feature with variance below a set threshold. For this paper, a $\chi^2$-test is utilized to select features. The $\chi^2$-test gives an indication of which features the target value is dependant of. 


\textit{Principal Component Analysis:} This involves identifying a set of linear combinations of original features that are orthogonal to each other. From the set, the first few combinations which explain the variance in the data above a certain threshold are chosen and used as features during the training.    

\textit{Autoencoders:} These can be seen as a nonlinear version of PCA. They are simply a neural network with a bottleneck layer. The dimension of the bottleneck layer is significantly smaller than the input dimensions. The input and output in a neural network are exactly the same. When trained the bottleneck layer compresses the information contained in the original data but with a significantly reduced dimension.

\subsection{Machine Learning Algorithms}
Several supervised and unsupervised algorithms are used in the current study. Within the supervised learning algorithms we use the following methods:

\textit{Artificial Neural Network (ANN):} It refers to a network of nodes, with connections carrying weight \cite{bala2017classification}. The network consists of one input layer, one or more hidden layers, and a final output layer. The learning part of this supervised learning algorithm is done by adjusting the weights and biases of the connections between the nodes. To learn non-linear patterns, each node is equipped with a non-linear activation function.

\textit{Support Vector Machine (SVM):} It is an ML method aiming to create a decision boundary between the classes in a dataset. The placement of this decision boundary is based on wanting the closest points in each of the classes to be as far away from the boundary as possible. 

\textit{Random Forest (RF):} It is a popular ML method, that is robust against overfitting. As the name implies, a RF consists of multiple decision trees, that operate as an ensemble. Each tree in the forest will come up with a classification, and the majority vote among the trees becomes the final output. 

\textit{$k$-Nearest-Neighbors ($k$NN):} It is a non-parametric classification method \cite{KNN}. In short, the algorithm classifies the input based on its distance to the $k$ nearest neighbors. The input is classified with the majority class among the  $k$ nearest neighbors.

The unsupervised algorithms which do not need any labels for clustering used in this work are as follows:

\textit{K-means Clustering:} It aims to partition observations into $K$ clusters. The assignment of observation to a cluster is done by measuring the distance from the observation to the cluster centroid.

\textit{Self Organised Map (SOM):} It is a sheet-like type of neural network, where the neurons compete with each other in the activations, and thus adaptively develop the ability to detect signal patterns \cite{kohonen1990self}. To begin with, each node in the node-grid is initialized with random vectors, with the same shape as the input data. To train the node-grid, also known as the codebook, random data points are presented individually. To calculate the similarity between the input and the nodes in the grid, the distance is calculated. Once all the distances are found, the winner, known as the Best Matching Unit, BMU, is the node closest in distance to the input. The BMU and a neighborhood around it, adapt to the input, based on a learning function. To stabilize the codebook over time, some decay function is given, which reduces the learning rate and the radius of the neighborhood for each iteration \cite{lakshminarayanan2020application}. 
\section{Data}
\label{section:data}
\begin{figure}
        \centering
        \includegraphics[width=\linewidth]{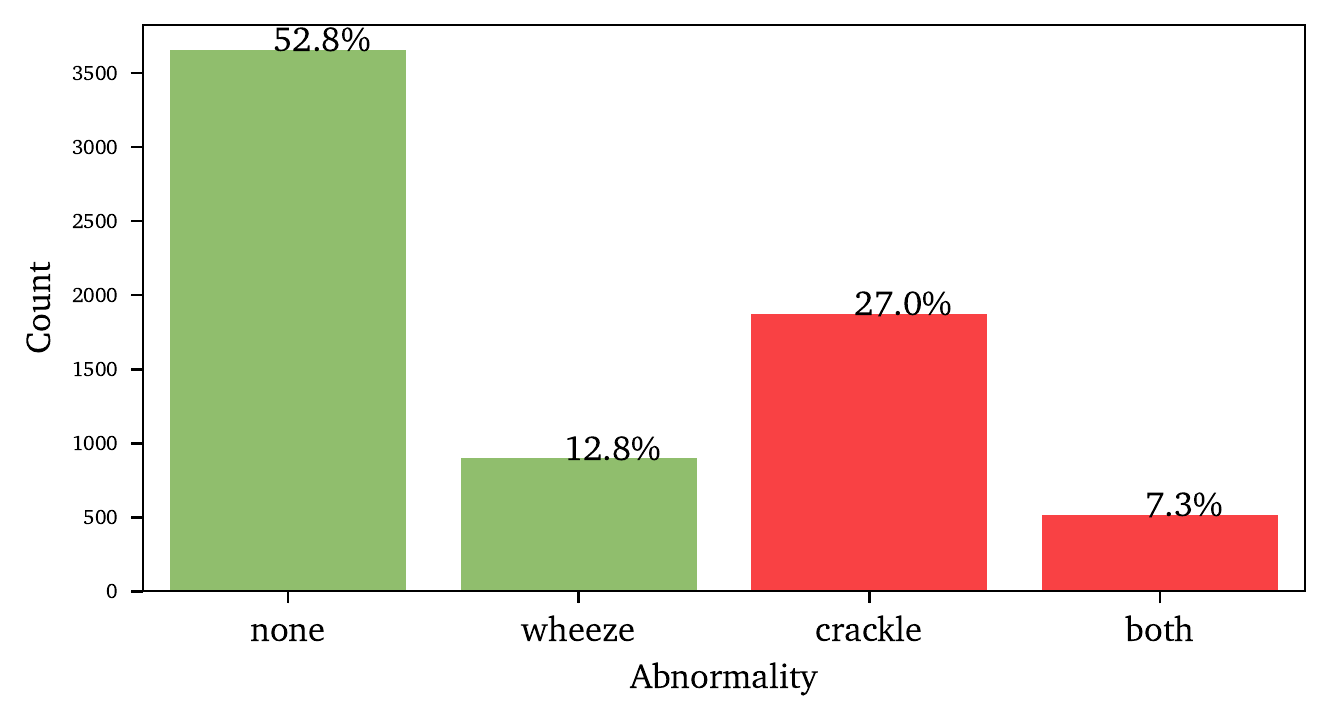}
        \caption{Histogram showing the distribution of the data and targets. \textbf{General dataset:} \textit{no-crackle =} none and wheeze, \textit{crackle =} crackle and both. \textbf{Pure dataset=} \textit{no-crackle = } none, \textit{crackle =} crackle}
        \label{fig:histoCrackleWheeze}
    \end{figure}

    The dataset used in this work, \cite{KaggleData}, was created as a scientific challenge at the International Conference on Biomedical Health Informatics - ICBHI 2017. We use this dataset because it makes the whole study reproducible. The dataset contains respiratory sounds, recorded at the School of Health Sciences, University of Aveiro, Portugal (ESSUA) and the Aristotle University of Thessaloniki, Greece (AUTH). The two universities used different types of recording equipment and to some extent different chest locations for the recordings. There are 920 audio files in the dataset, as well as 920 annotation files, taken from 126 patients. The annotation files identify the start and end of the breathing cycles within the audio files, where the breathing cycles are marked as crackle, wheeze, or normal. For exploration, two separate baseline datasets were obtained. Firstly, a more general dataset was created, where the data labeled as `wheeze' was concatenated with the data labeled as `none' and relabelled as `no-crackle'. The data labeled as `both' was concatenated with `crackle' to create the label `crackle'. Figure \ref{fig:histoCrackleWheeze} gives a distribution of the labelled dataset.
    
    \subsection{Data Preprocessing} \label{sec: pre-process}
    It is known that LS exhibits non-periodicity and low frequency in the range $50-2500 Hz$ \cite{reichert2008analysis}. Also looking through the FFT of the data (Figure \ref{fig:Frequency_domain_5_Samples}), one can observe that this assumption is valid. Hence, as pointed out in \cite{10.1007/978-981-10-7419-6_7}, it is considered logical to downsample the LS. In the mentioned paper the authors downsampled the data to 4kHz, however since there are some differences in literature, as to what is the maximum frequency observed in the LS, a frequency of $8kHz$ is chosen in this paper. This step of downsampling allows for testing out decomposition methods, such as EMD and EEMD at a larger scale, because of the reduced computation time. 
    
    \begin{figure}
        \centering
        \begin{center}
            \includegraphics[width=\linewidth]{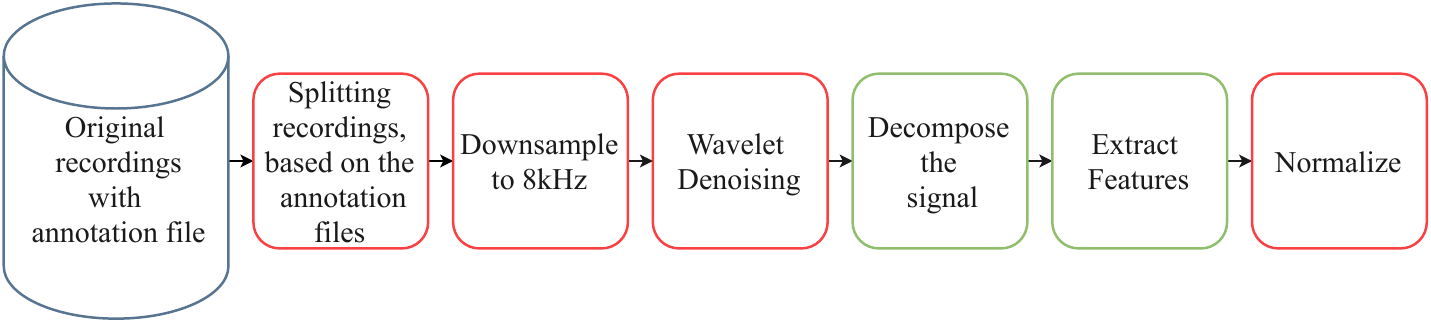}
            \caption{The proposed procedure for pre processing}
            \label{fig:preProcessing}
        \end{center}
    \end{figure}
    
\begin{figure}%
        \centering
        \subfloat[\centering Scatterplot]{{\includegraphics[width=\linewidth]{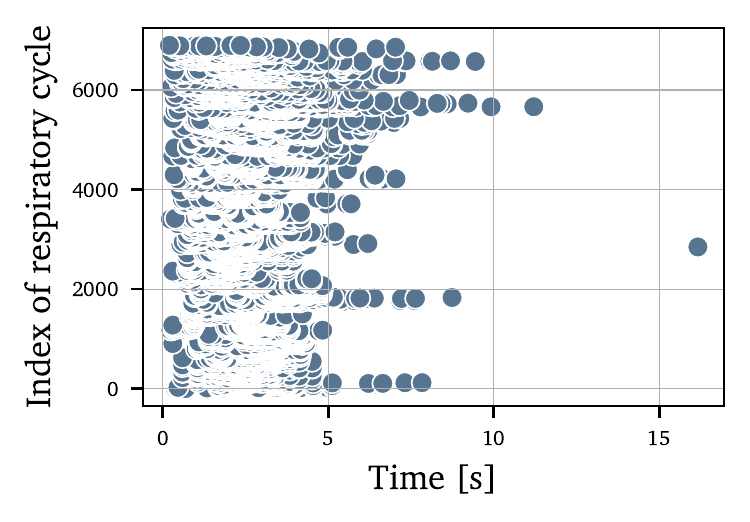} }}%
        \qquad
        \subfloat[\centering Histogram]{{\includegraphics[width=\linewidth]{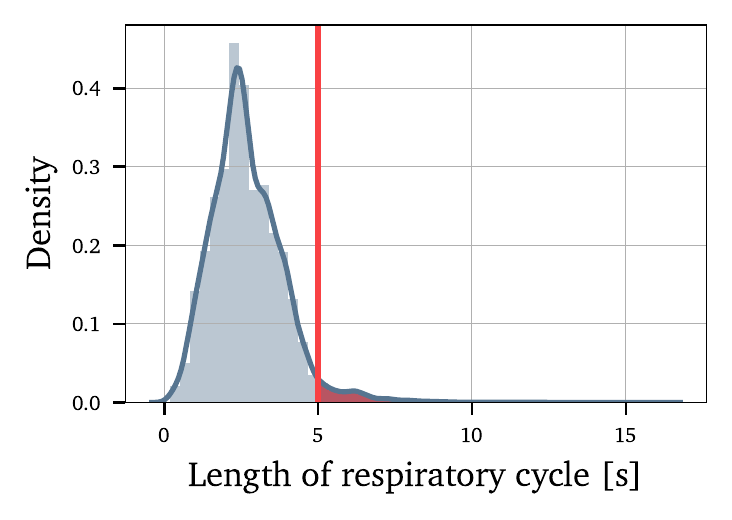} }}%
        \caption{Plots of the lengths of the respiratory cycles in the dataset}%
        \label{fig:scatter and hist len}%
    \end{figure}

    All the audio files are split into their labeled respiratory cycles, in order to specify the region of interest. The lengths of the respiratory cycles varies, depending on the shortness of the breath of the patient being recorded, this can be seen in Figure \ref{fig:scatter and hist len}. For the reproduction of the Convolutional Neural Network (CNN) results it is desirable to have equal lengths for all the timeseries, hence some fixed signal length needs to be found. From inspecting the scatter plot in Figure \ref{fig:scatter and hist len}, one can notice the presence of outliers in the dataset. In order to filter out these outliers, one can decide on a set length for all the slices. By setting an upper limit of 5 seconds, the distribution plot in Figure \ref{fig:scatter and hist len} shows that 96.6 \% of all the data is shorter than this upper limit. From the research conducted, it is known that respiratory cycles usually last for about 5 seconds, hence the upper limit given is considered valid. 
    
\begin{figure}
        \begin{center}
            \includegraphics[width=\linewidth]{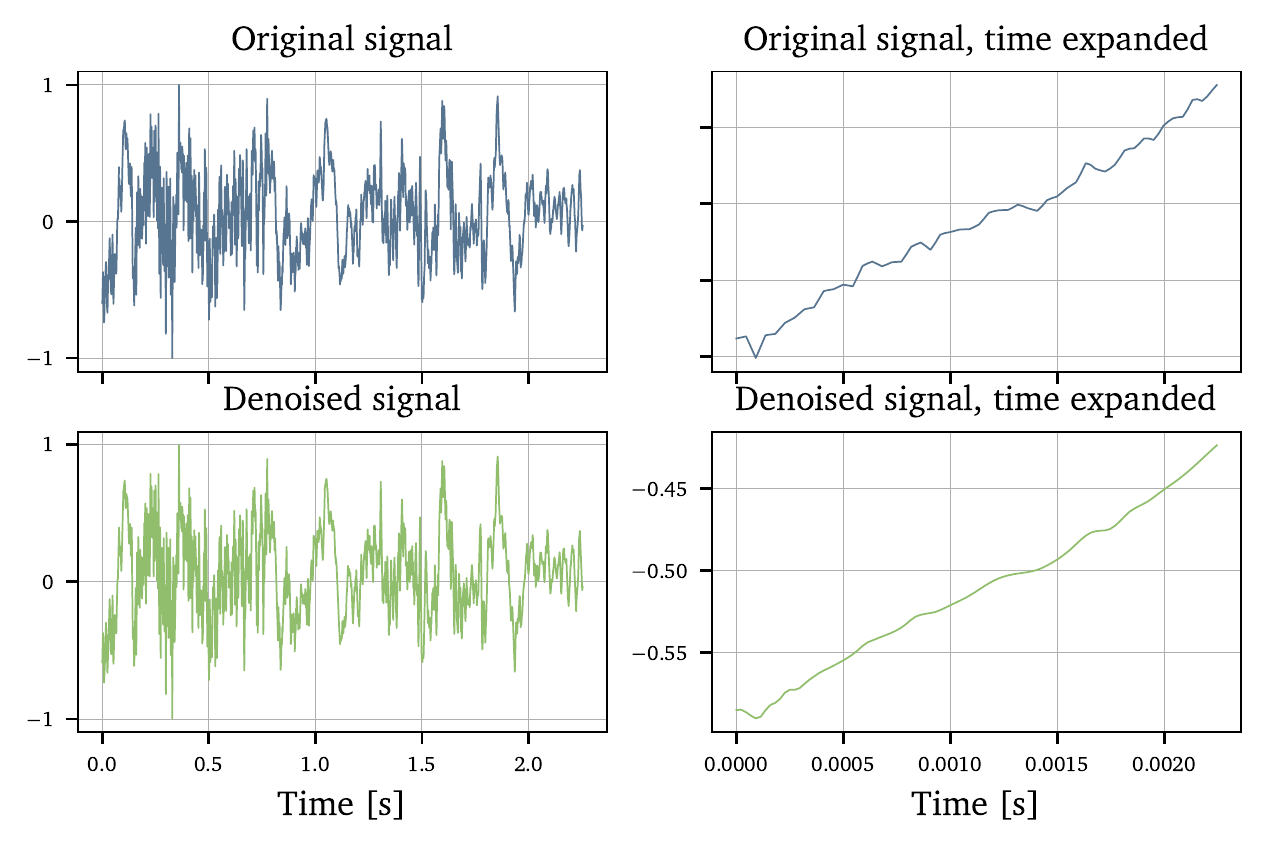}
            \caption{The original LS (upper) compared to the denoised LS (lower). The left side of the plot shows the time expanded signal, to clearly see the difference}
            \label{fig:denoised}
        \end{center}
    \end{figure}

    After clipping the signals to the appropriate length, the denoising method proposed in \cite{bahoura1998respiratory} is performed on the clipped signals. In Figure \ref{fig:denoised} a representative respiratory cycle (upper) is compared to the denoised version (lower).
    
    \subsection{Data Exploration}
    The data presented has audio-clips containing individual respiratory cycles as input, before preprocessing. The label values for the respiratory cycles are either no-crackle or crackle.
    
    Other variables included in the process for classification are determined following the procedure of preprocessing the input signal. The variables intended to be the input of the actual classifier are the ones processed through the steps of preprocessing, feature extraction, and feature selection. 
    
\begin{figure}
        \begin{center}
            \includegraphics[width=\linewidth]{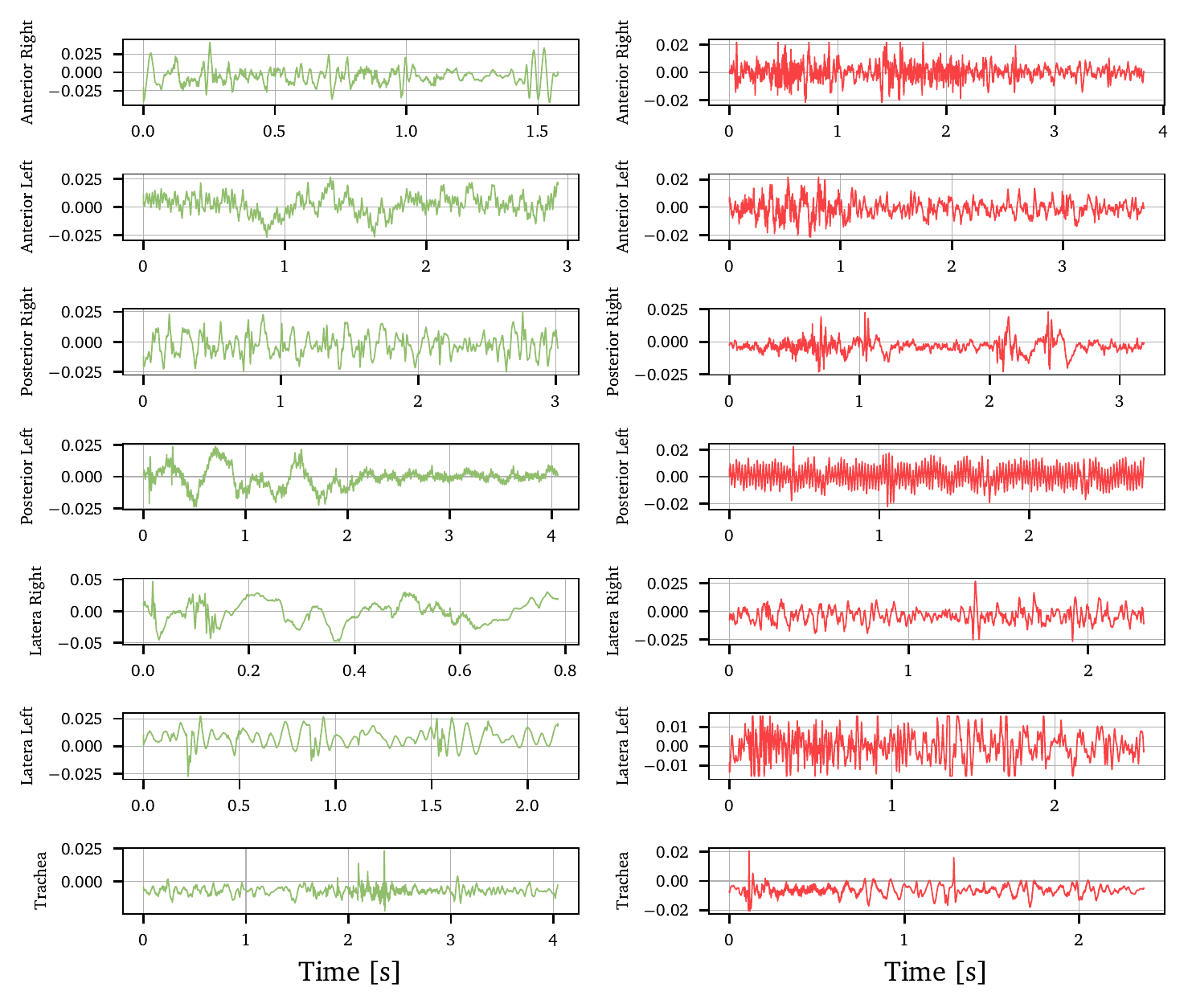}
            \caption{Samples of no crackle (left) and of crackle (right) in time domain, based on recording location. Normalized audio. }
            \label{fig:Time_domain_5_samples}
        \end{center}
    \end{figure}

\begin{figure}
        \begin{center}
            \includegraphics[width=\linewidth]{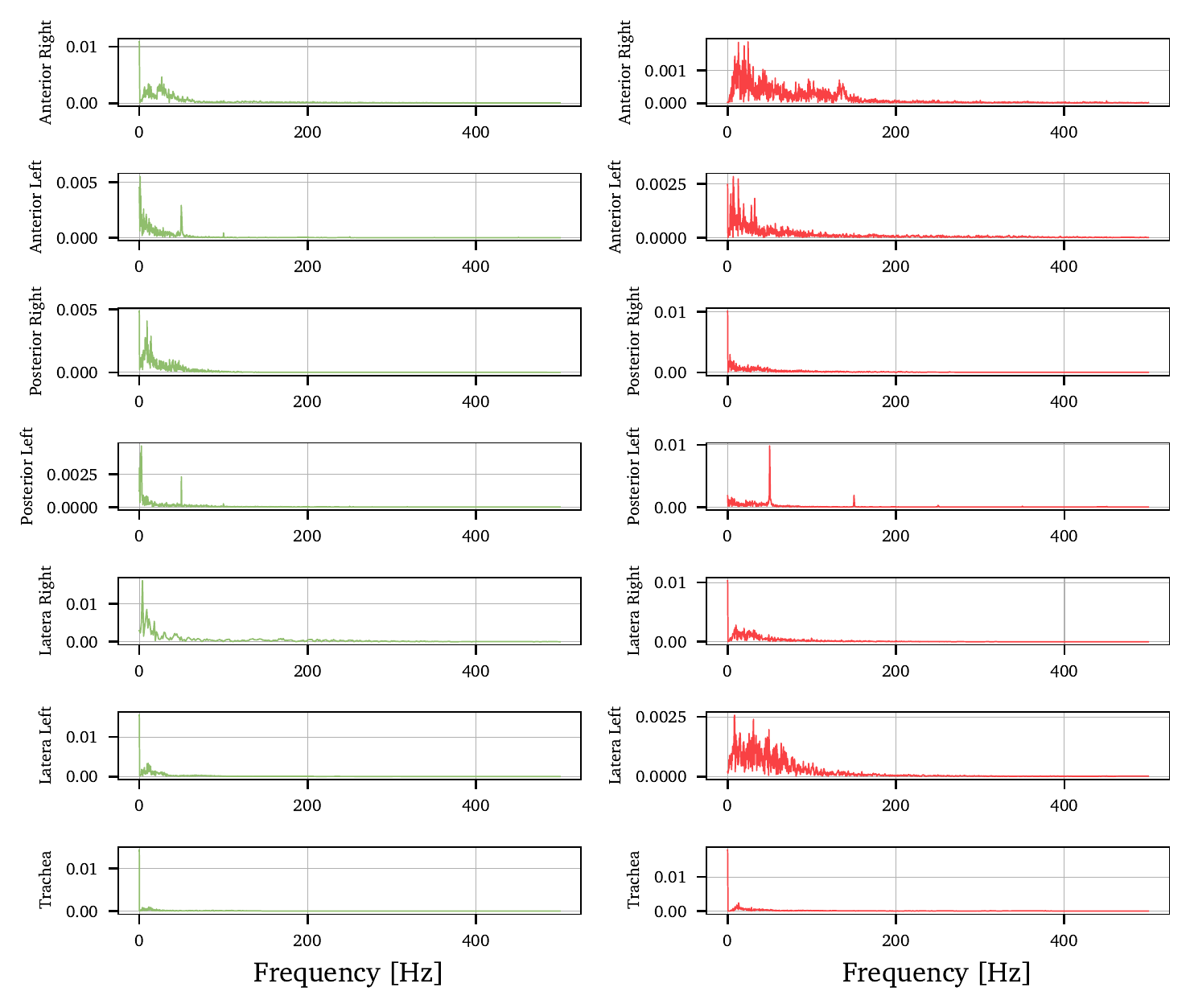}
            \caption{Samples of no crackle (left) and of crackle (right) in frequency domain, based on recording location. Normalized audio. }
            \label{fig:Frequency_domain_5_Samples}
        \end{center}
    \end{figure}

    To get some idea of what kind of signals one is dealing with, several plots for comparison are displayed in Figures \ref{fig:Time_domain_5_samples} and \ref{fig:Frequency_domain_5_Samples}. In the figures, a random sample of respiratory cycles marked as crackle or no-crackle is selected, based on the recording location. To simplify the comparison, since the LS is quite varying over the entire dataset, the audio clips are normalized for all the plots. 
    
    In Figure \ref{fig:Time_domain_5_samples} the respiratory cycles marked as no-crackle (left) and crackle (right) are compared in the time domain. With a mere visual inspection these figures one could hardly distinguish the two classes and hence they are again presented in the Fourier domain (Figure \ref{fig:Frequency_domain_5_Samples}). From this figure it is clear that the signal with crackle have a dominance of the higher frequency components. 
    
\section{Set-up}
\label{section:setup}
\begin{figure}
        \begin{center}
            \includegraphics[width=\linewidth]{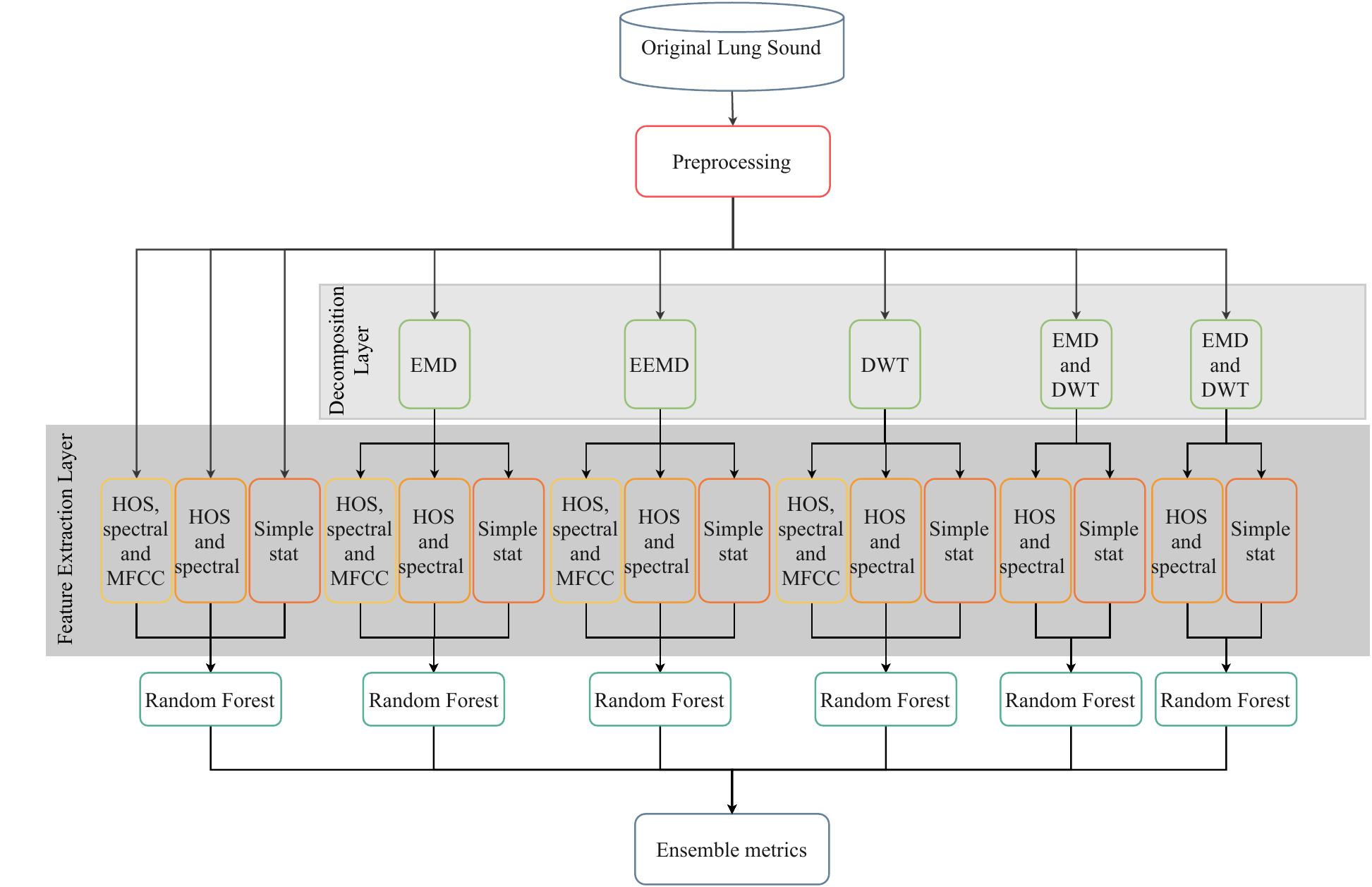}
            \caption{Proposed procedure for comparing different decomposition and feature extraction methods, to decide on the most representative feature set. }
            \label{fig:processing procedure}
        \end{center}
    \end{figure}

    Classification reports is to be collected to compare the performance of the decomposition method, along with the feature extraction procedure. To avoid inspecting all combinations of feature sets, decomposition methods, and classifiers, a RF algorithm is utilized to decide on the most representative feature set, according to Figure \ref{fig:processing procedure}. After the best combinations of decomposition technique and feature sets are decided, these are employed to explore the other classifiers. The total workflow is visualized in Figure \ref{fig:workflow}. 
    
    \subsection{Decomposition and Feature Extraction} \label{sec: feature and decomp res}
    \begin{figure*}
         \centering
         \begin{subfigure}[b]{0.47\textwidth}
             \centering
             \includegraphics[width=\textwidth]{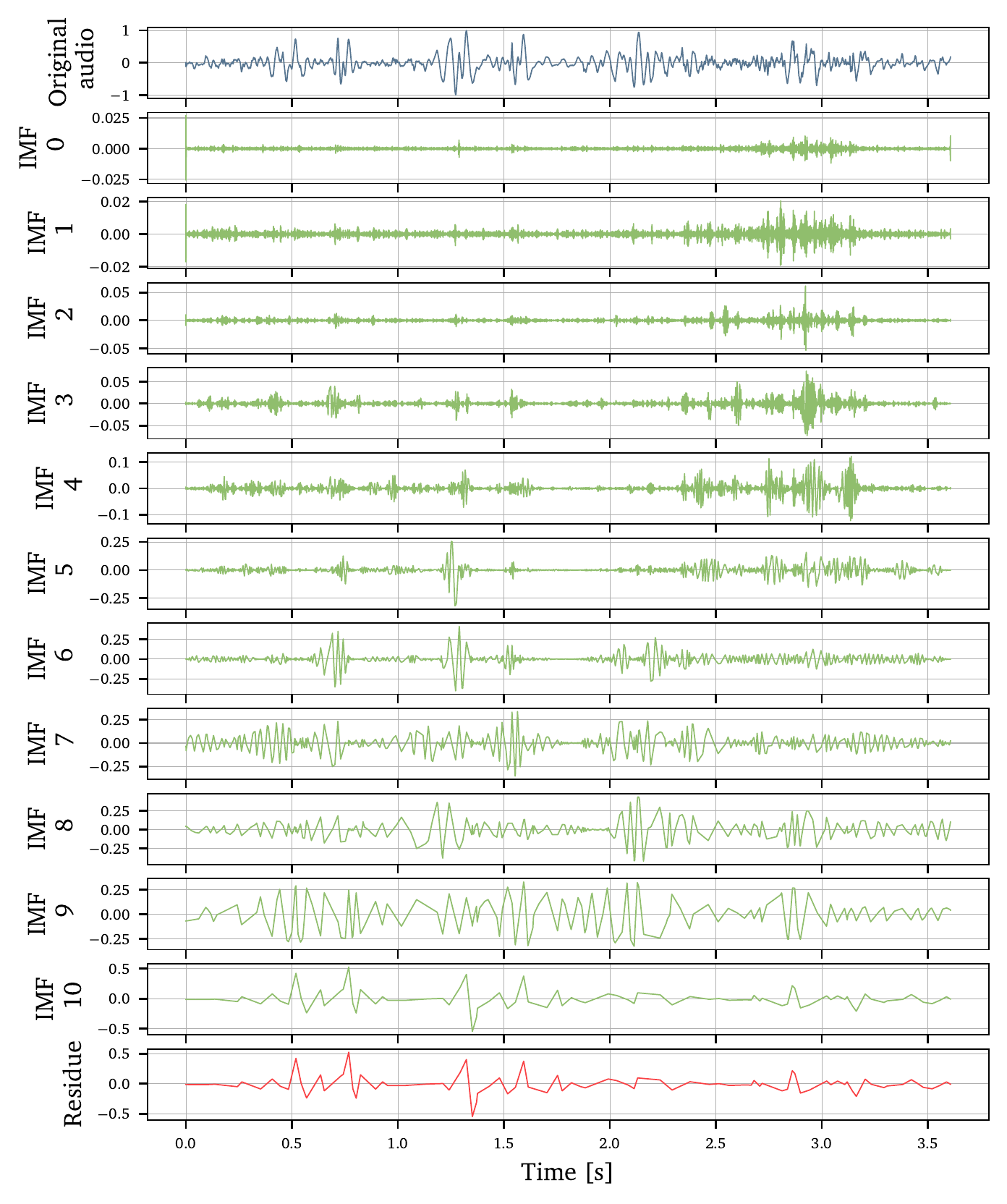}
             \caption{EMD}
             \label{fig: EMD}
         \end{subfigure}
         \hfill
         \begin{subfigure}[b]{0.47\textwidth}
             \centering
             \includegraphics[width=\textwidth]{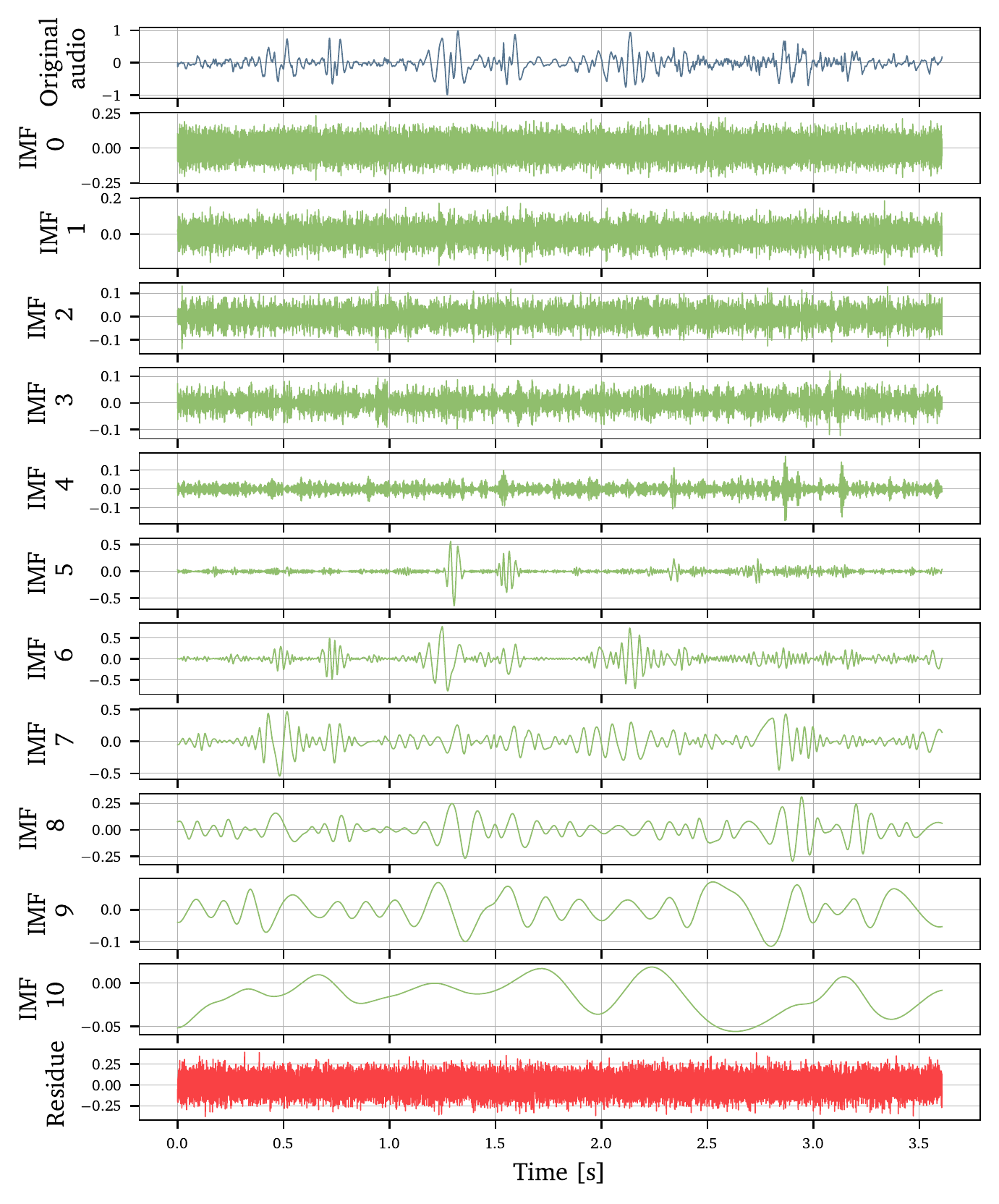}
             \caption{EEMD}
             \label{fig: EEMD}
         \end{subfigure}
         \caption{Representative plots of the IMF's}
    \end{figure*}
    
    It is desirable to derive a compact representation of the LS, in order to make efficient and accurate classifications. Feature sets based on the feature extraction methods explained in Section \ref{sec: Theory Feature Extraction}, are applied to the decomposed signal.
    
    Furthermore to reduce the dimension of the \textit{spectral} and \textit{MFCC} features, and assure equal feature-vector sizes for breathing cycles of varying length, 6 statistical features are computed for each of the feature vectors presented. The statistical features extracted from the spectral- and MFCC- feature vectors are; mean, standard deviation, skewness, maximum, median, and min.
    
    Six distinctive decompositions are created, based on EMD, EEMD, DWT, and EMD/EEMD in combination with DWT. Along with the six decompositions, three feature sets are explored. The first feature set, marked as the \textit{simple}-feature set, consists of the simple features presented in Section \ref{sec: Theory Feature Extraction}. A more complex feature set, labeled \textit{HOS and spectral}, is attained extracting the simple features, along with the higher-order statistical and spectral features. Finally, the largest feature set, contains both the \textit{simple}- and  \textit{HOS and spectral}-features, along with the \textit{MFCC}-features. A summary of the explained steps for combining and comparing feature sets with decomposition methods is displayed in Figure \ref{fig:processing procedure}. 
    
    For EMD and EEMD a maximum of 10 IMF's and 10 sifting iterations were set. These two criteria ensure fast computations, as well as avoiding over sifting. From the knowledge that the lower level modes contain higher frequencies, and crackles contain mostly higher frequencies, only the first 5 modes are selected, after running both EMD and EEMD. 
    
    Expectedly, EMD came with the drawback of mode-mixing, which is displayed in the Figure \ref{fig: EMD},  hence there is no guarantee that the modes selected would represent the same characteristics. EEMD solves the mode-mixing issue, even when only applying two ensembles, which can be seen in Figure \ref{fig: EEMD}. Because of the computational effort, a higher number of ensembles are not considered. Additional ensembles could lead to a more stable decomposition, ensuring that the IMF's satisfy the IMF criteria. 
    
    DWT is used by a multitude of papers \cite{kandaswamy2004neural, ulukaya2017overcomplete, 7591542}, to decompose the LS, thus, this decomposition approach is also investigated in the current work. The signal is decomposed into 10 levels of detail coefficients. The wavelet 'db8' is used, as proposed in \cite{kandaswamy2004neural}. From the theory behind DWT, it is known that for each level of decomposition the signal is half band filtered and down-sampled by 2. Since the signal is also down-sampled before performing DWT the highest level of approximation coefficients represents the frequency range $[0 , f_s / 2^J]$, where $J$ is 10 in this case. After down-sampling to $8kHz$, this correspond to $[0 Hz , 7.8 Hz]$, which represent frequencies of no interest, in regards to classifying crackles, hence these approximated coefficients are discarded.
    
    It is assumed that the ability to discriminate between no-crackle and crackle, is not weakened by using a combination of EMD/EEMD and DWT to decompose the signal. The IMF's resulting from EMD and EEMD, is thus further decomposed using DWT. Feature extraction, based on the three feature sets explained previously, is performed on the detail coefficients of the decomposed IMF's. Again, only the five first IMF's are used, also no MFCC features are extracted. These two adjustments are the results of limitations in storage. The choice of only including 5 levels of detail coefficients is thought to be valid because the highest level of the detailed coefficients represent frequencies with a minimum of $250Hz$. From the knowledge that crackles lies between $200-2500Hz$, this minimum is accepted. 
    
    Once the features are extracted, using one of the proposed methods for feature extraction, further preprocessing is needed. To make the optimization algorithm run faster the extracted features are normalized to the range $[0,1]$. 
    
    A short summary of the steps involved to extract the features from the raw audio recordings is displayed in the Figure \ref{fig:preProcessing}.
    
    \subsection{Feature Selection}
    In some of the discussed combinations of the decompositions and feature extractions, the dimension of the input becomes as large as 968 features, many of them being redundant. These redundant features can make the model unnecessarily complex and hence the feature selection methods are used to remove the redundancies. For feature selection three approaches mentioned in Section \ref{subsec:featureselection}; filtering, PCA and Autoencoder are explored. For feature selection based on filtering a $\chi^2$-test is employed to evaluate the relevance of the features.  
    
    \begin{table*}
    \centering
    \caption{Hyperparameters found for supervised classification, using 5-fold cross validation. Only displaying parameters set to something other than the default} \label{tab: hyperparam}
   \begin{tabular}{p{1cm} p{2cm} p{1.5cm} p{2.5cm} p{1.5cm} p{2.5cm} p{1.5cm} p{1.5cm}}
    \toprule
    Classifier & \makecell{Hyper- \\ parameter}   &\multicolumn{6}{c}{Decomposition}\\
    \midrule
    \multicolumn{2}{c}{}   &       \makecell{No \\ decomposition}      & EMD    & EEMD   &  DWT & EMD \& DWT & EEMD \& DWT \\
    \midrule
    RF & {}   &   {}              & {}     &    {}  &   {} &    {}      &  {}          \\
    {}  & Max depth                     &    12             & 12     & 8    & 12   & 10  & 8 \\ 
    {}  & \makecell{N- \\ estimators}                  &    {}             & 300     & 300    & {}   & {}  & 300 \\ 
    
    ANN & {}   &  {}     & {}     &    {}  &   {} &    {}      &  {}          \\
    {}  & \makecell{Hidden- \\ layer size}             &    [1000,500]   & [1000,500, 250] & [1000,500] &  [1000,500, 250] &  [1000,500] & [1000,500] \\ 
    {}  & \makecell{Learning- \\  rate  ($\alpha$) }              &    {}             &  {}     & 0.05   & 0.05   & {}  & 0.05 \\ 
    
    SVM &  {}   &  {}              & {}     &    {}  &   {} &    {}      &  {}          \\
    {}  & C                     &    10             & 1     & 1    & 10   & 1  & 1 \\ 
    {}  & $\gamma$                  &   1            & 0.1     & 0.1    & 0.1   & 0.1  & 0.1 \\ 
    $k$-NN & {}   &  {}              & {}     &    {}  &   {} &    {}      &  {}          \\
    {} & Leaf-size & 1 & 1 & 1 & 1 & 1 & 1 \\
    {} & \makecell{N- \\ neighbours}   & 1 & 2 & 1 & 1 & 2 & 1 \\
    {} & p & 1 & 1 & 1 & 1 & 1 & 1 \\
    \bottomrule
    \end{tabular}
    \end{table*}
\section{Results}
\label{section:results}
    In this section the results are presented and discussed, in order to make an educated decision on what procedure works best for analyzing and detecting anomalies in LS data.
    
    \begin{figure}
        \centering
        \includegraphics{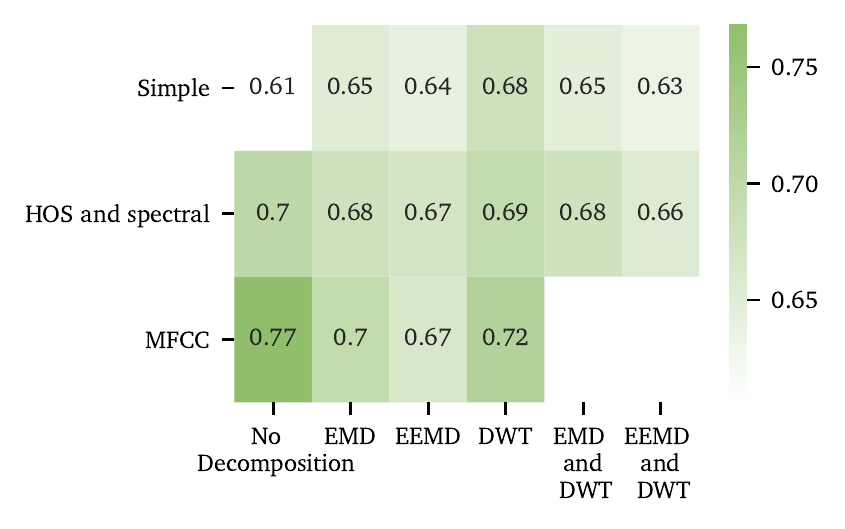}
        \caption{Comparing the feature extraction and decomposition combinations, based on f1-score. Here a RF is used for classification. }
        \label{fig:comparing_feature_extractio_f1_score}
    \end{figure}
    
    \subsection{Classification using the full feature set}
    \label{subsec:classification_using_the_full_feature_set}
        To decide on the best combination of decomposition, feature set and classifier, several combinations were explored. Figure \ref{fig:comparing_feature_extractio_f1_score} summarizes the result of all the decomposition approaches, along with the three chosen feature extractors. The dimension of the feature set could be as large as 980 features. Here a simple RF classifier was utilized to compare the performance of the decomposition and feature extraction combinations. Since from the figure it is clear that the MFCC consistently provides the best feature extraction, it is chosen as the preferred method for feature extraction throughout the work.
        
        \begin{figure}
            \begin{center}
                \includegraphics[width = \linewidth]{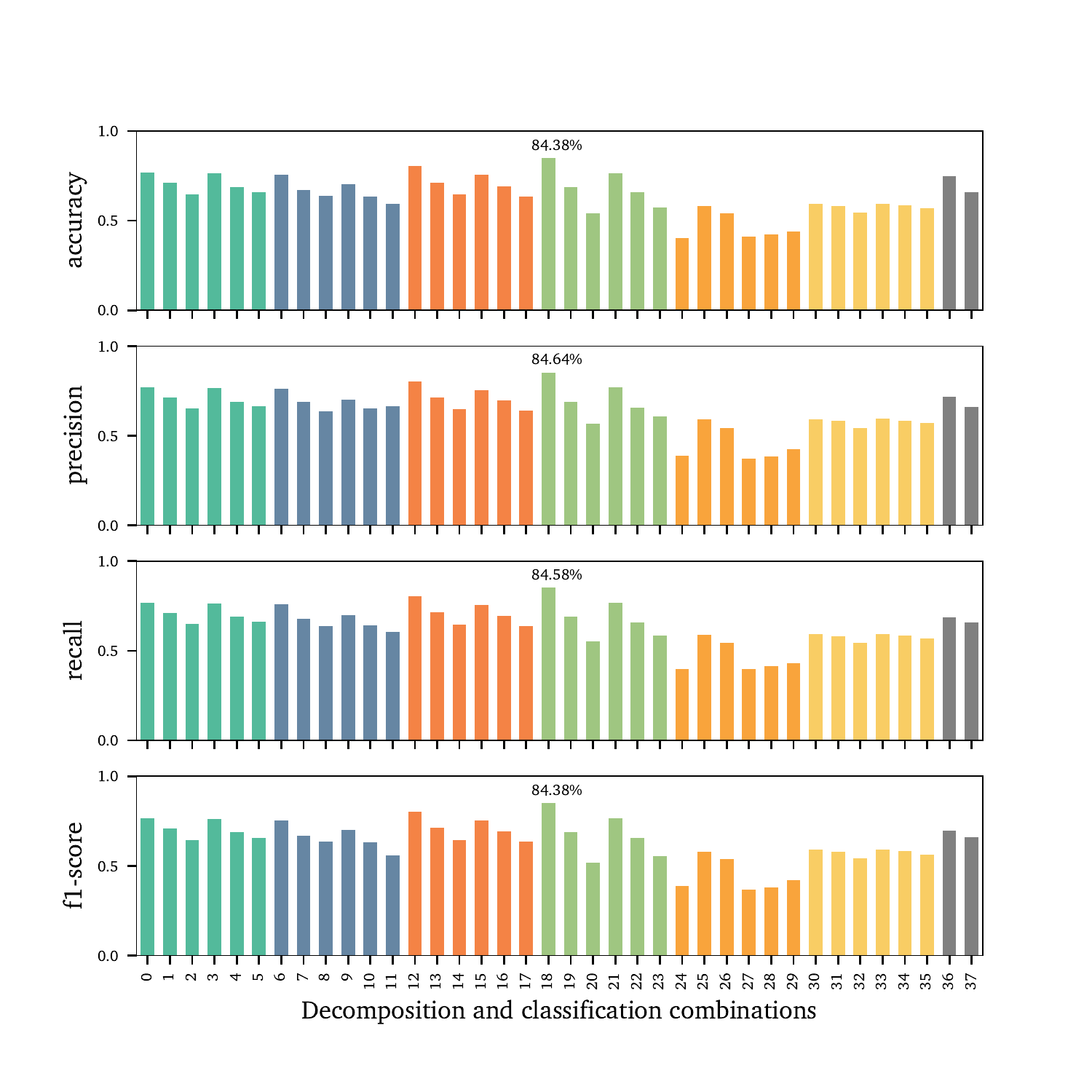}
                \caption{Bar plot of accuracy, precision, recall and f1-score divided into sub-figures. Mapping from number on the x axis to decomposition and classifier is as follows: \textbf{0}:No decomposition,RF \textbf{1}:EMD,RF, \textbf{2}:EEMD,RF \textbf{3}:DWT,RF \textbf{4}:EMD,DWT,RF \textbf{5}:EEMD ,DWT,RF \textbf{6}:No decomposition,ANN \textbf{7}:EMD,ANN \textbf{8}:EEMD, ANN \textbf{9}:DWT,ANN \textbf{10}:EMD,DWT,ANN \textbf{11}:EEMD,DWT,ANN \textbf{12}:No decomposition,SVM \textbf{13}:EMD,SVM \textbf{14}:EEMD,SVM \textbf{15}:DWT,SVM \textbf{16}:EMD,DWT,SVM \textbf{17}:EEMD,DWT,SVM \textbf{18}:No decomposition,k-NN \textbf{19}:EMD,k-NN \textbf{20}:EEMD, k-NN \textbf{21}:DWT,k-NN \textbf{22}:EMD,DWT,k-NN \textbf{23}:EEMD,DWT,k-NN, \textbf{24}:No decomposition,k-means \textbf{25}:EMD,k-means \textbf{26}:EEMD,k-means \textbf{27}:DWT,k-means \textbf{28}:EMD,DWT,k-means, \textbf{29}:EEMD,DWT,k-means \textbf{30}:No decomposition,SOM \textbf{31}:EMD,SOM \textbf{32}:EEMD,SOM \textbf{33}:DWT,SOM \textbf{34}:EMD,DWT,SOM \textbf{35}:EEMD,DWT,SOM \textbf{36}:CNN \textbf{37}:Recreated Paper]}
                \label{fig:all_metrics_in_subfigures}
            \end{center}
        \end{figure}
        
        \begin{figure}
            \centering
            \includegraphics{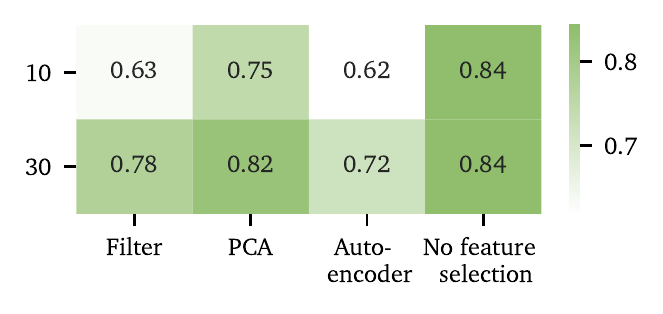}
            \caption{Heatmap comparing feature selection approaches, compared to no feature selection, in terms of accuracy. The best approach, being $k$-NN in combination with no decomposition, is used for classification. }
            \label{fig:featureSelectionHeatMap}
        \end{figure}
        \begin{figure}
            \centering
            \includegraphics[width = \linewidth]{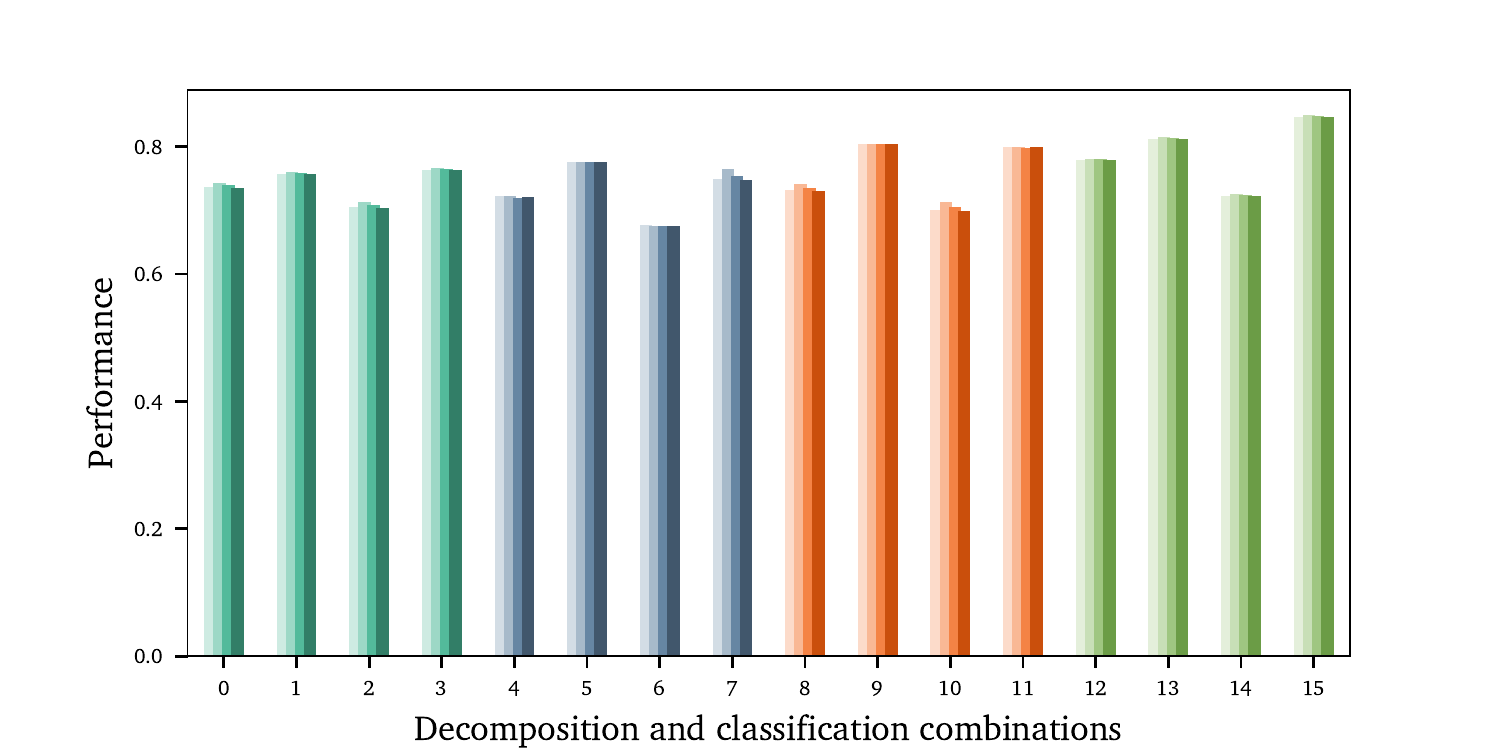}
            \caption{The gradient, when going from light-shade to dark-shade, represents: accuracy, precision, recall and f1-score respectively. Each section of bars are connected to the following combinations: \textbf{0}: RF,Filtering,\textbf{1}:RF, PCA, \textbf{2}:RF,Autoencoder, \textbf{3}:RF,No feature selection, \textbf{4}: ANN,Filtering, \textbf{5}:ANN,PCA, \textbf{6}:ANN,Autoencoder, \textbf{7}:ANN,No feature selection, \textbf{8}:SVM,Filtering, \textbf{9}:SVM,PCA, \textbf{10}:SVM, Autoencoder, \textbf{11}:SVM,No feature selection, \textbf{12}:KNN,Filtering, \textbf{13}:KNN,PCA, \textbf{14}:KNN,Autoencoder, \textbf{15}:KNN, No feature selection}
            \label{fig:No_decomp:_feature_selection_classifier_combinations}
        \end{figure}
        
        \begin{figure}
            \centering
            \includegraphics[width = \linewidth]{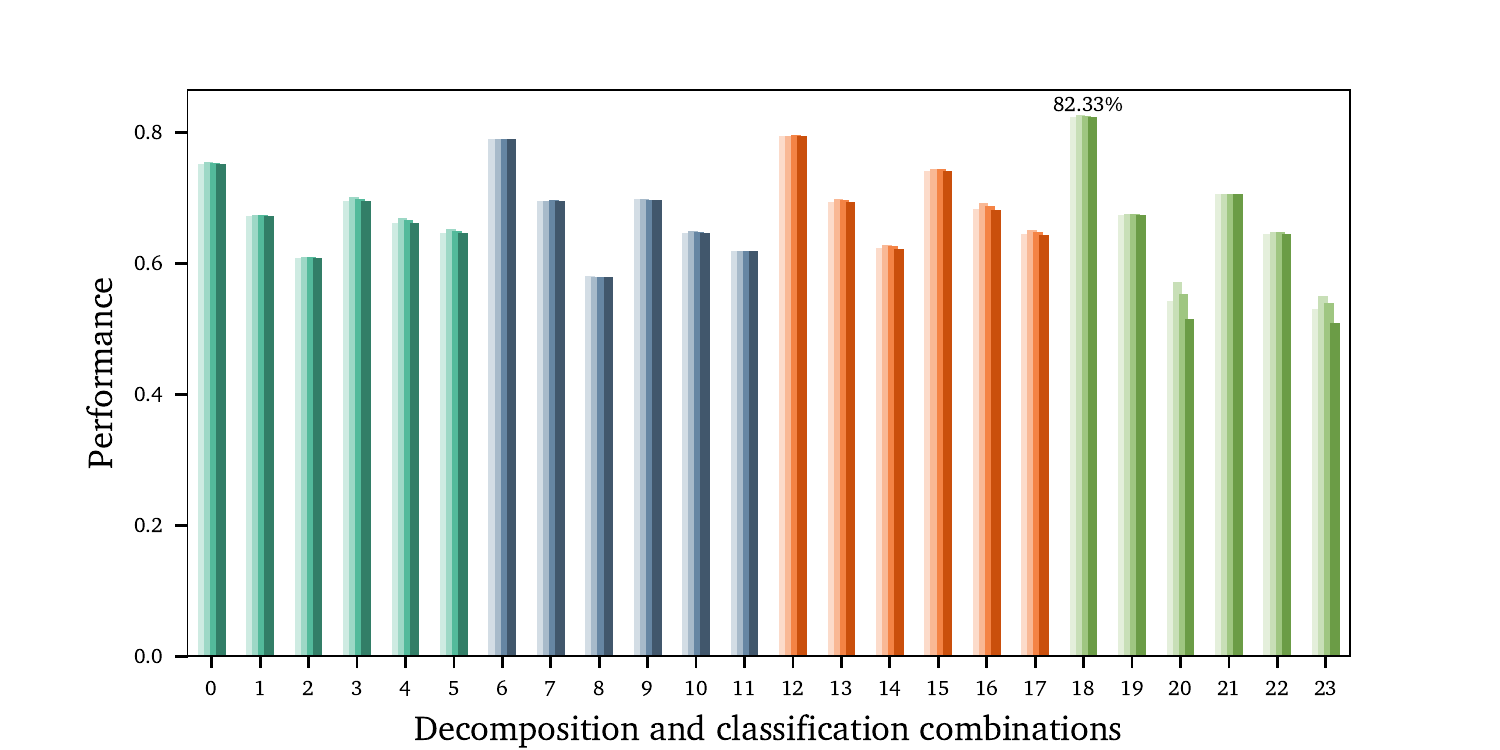}
            \caption{The gradient, when going from light-shade to dark-shade, represents: accuracy, precision, recall and f1-score respectively. Each section of bars are connected to the following combinations: \textbf{0}:RF,No Decomposition, \textbf{1}:RF,EMD, \textbf{2}:RF,EEMD, \textbf{3}:RF,DWT, \textbf{4}:RF,EMD,DWT, \textbf{5}:RF,EEMD,DWT, \textbf{6}:ANN,No Decomposition, \textbf{7}:ANN,EMD, \textbf{8}:ANN,EEMD, \textbf{9}:ANN,DWT, \textbf{10}:ANN,EMD,DWT, \textbf{11}:ANN,EEMD,DWT, \textbf{12}:SVM,No Decomposition, \textbf{13}:SVM,EMD, \textbf{14}: SVM,EEMD, \textbf{15}:SVM,DWT, \textbf{16}:SVM, EMD,DWT, \textbf{17}:SVM,EEMD,DWT, \textbf{18}: K-NN,No Decomposition, \textbf{19}:K-NN,EMD, \textbf{20}:K-NN, EEMD, \textbf{21}:K-NN,DWT, \textbf{22}:K-NN,EMD,DWT, \textbf{23}:K-NN,EEMD,DWT}
            \label{fig:PCA30:Decomposition_and_classifier_combinations}
        \end{figure}
        Figure \ref{fig:all_metrics_in_subfigures}, presents the different metrics after combining the various decomposition approaches with the different classifiers. To attain the results in Figure \ref{fig:all_metrics_in_subfigures}, an exhaustive search for hyperparameters was performed, along with 5-fold cross-validation, to ensure unbiased tuning. The resulting hyperparameters, for the various decomposition and classifier combinations, are displayed in the Table \ref{tab: hyperparam}. The Figure \ref{fig:all_metrics_in_subfigures} shows a clear tendency in all classifiers; the features without decomposition, proved most effective for the classification. Furthermore, $k$-NN outperformed all other classifiers.
        
        What is surprising is that EMD and EEMD which are very popular decomposition techniques for extracting features from the ECG and EEK data did not improve the classification. In fact the inclusion of the features obtained from the decomposition degraded the classification. One reason could be the fact that including these decomposition techniques increases the number of the features drastically resulting in classification models that could be overly complex and hence prone to overfitting.  
        
    \subsection{Classification using selected features}
    \label{subsec:classification_using_selected_features}
        As mentioned earlier, the number of features used in the classification in the previous section could be upto 980 depending on the decomposition and feature extractor employed. In this section we explore if the number of features could be reduced without compromising on the accuracy. To compare the performance of the feature engineering and selection approaches discussed in the Section \ref{section:theory}, a selection of 30 features were investigated. Figure \ref{fig:featureSelectionHeatMap} was obtained by running the three discussed feature selection methods, along with the best performing decomposition, feature extractor, and classifier combinations. It is clear from the Figure \ref{fig:featureSelectionHeatMap}, that the feature selection based on PCA gave the best accuracy. 
        
        To further inspect the effect of feature selection, Figures \ref{fig:No_decomp:_feature_selection_classifier_combinations} was plotted. Figures \ref{fig:No_decomp:_feature_selection_classifier_combinations} represents the metrics obtained when different combinations of feature selection and classifiers are explored. Here 30 most important features are utilized. From the figure it can be concluded that classification based on the 30 most important PCs obtained using PCA gave the best performance which was comparable to the performance of the classifiers run on the full feature set (~980).
        
        To make a more balanced comparison the the result earlier presented in Figure\ref{fig:all_metrics_in_subfigures} is reproduced but this time by considering only the 30 most important PCs obtained from the PCA. The result is presented in the Figure \ref{fig:PCA30:Decomposition_and_classifier_combinations}. It can be observed in Figure \ref{fig:No_decomp:_feature_selection_classifier_combinations}, that although there is a significant reduction in the number of the features (980 to 30), there is no noticeable drop in the performance. In fact for some classifiers, the performance actually improves.

    \subsection{Comparison to Previous Works}
    \label{subsec:comparison_to_previous_works}
        The two last bars in the Figure \ref{fig:all_metrics_in_subfigures}, correspond to the results presented in the paper \cite{10.1007/978-981-10-7419-6_8}, that had promising results in the conference challenge, for which the dataset was created. Both the results from the Kaggle implementation of the CNN, and the feature extraction based on resonance, and classification using an SVM from \cite{10.1007/978-981-10-7419-6_8}, seemed to give weaker results than the several of our proposed approaches (Figure \ref{fig:all_metrics_in_subfigures}).  
\section{Conclusions}
\label{section:conclusions}
    In this study, various methods for extracting relevant features from the LS recordings are explored. Through a comprehensive study of the decomposition methods, feature extraction, engineering and selection techniques, different supervised and unsupervised classifiers, the most promising combination was found to be MFCC features from a LS recording without decomposition. For classification, $k$-NN provided the best accuracy of 84.38\%. The major conclusions from the study can be summarised as: 
    \begin{itemize}
        \item The decomposition techniques like EMD and EEMD which are highly effective in extracting features from EEG and ECG are not very effective in improving the discrimination of LS between crackle and no-crackle.  
        \item $k$-NN proved to be the most accurate classifier, in combination with the dataset without any decomposition. 
        \item Compared to the previous works, employing the same dataset, the proposed approach for classifying between crackle and no-crackle seems more successful. 
        \item Application of various feature selection methods demonstrate that one can get similar classification accuracy with significantly reduced number of features. 
    \end{itemize}
    
    The dataset used for training and testing is composed of datasets from two different institutions which used slightly different techniques for data acquisition. This explains why we could not achieve comparable performance reported in other papers  \cite{kandaswamy2004neural, ulukaya2017overcomplete} which utilize another dataset. It is also worth mentioning that LS signals are characterized by non-periodicity and low frequencies in the range of $50-2500Hz$ \cite{reichert2008analysis}. Heart sound (HS) are also within the frequency range of $20-600 Hz$, with most of the frequency components falling in the range $20-150Hz$ \cite{mondal2017enhancement}. The two signals thus interfere with each other, necessitating the need for separating them before LS could be analysed. In previous works, \cite{10.1007/978-981-10-7419-6_7} and \cite{10.1007/978-981-10-7419-6_8}, where classification is performed on the same dataset as utilized in this project, the preprocessing step of minimizing the influence from the HS is performed. Several filters were proposed for the task. However, as stated, there is a significant overlap between the HS and the LS, hence simply filtering based on the frequencies, would discard a lot of valuable information about the LS. Independent Component Analysis which are very effective in separating non-Gaussian and statistically different signals can be applied to accomplish a better separation of the two sounds however, that will require all the data collection to be repeated with at least two measurement devices. 
    
    Towards the end we would like to stress that the approach utilized in this work can be applied to analyze and detect anomalies in any timeseries data, not necessarily biomedical signals. 
    
\section*{Acknowledgments}
This work was supported by the Research Council of Norway through the EXAIGON project, project number 304843


\bibliographystyle{unsrtnat}
\bibliography{ref}
\vskip3pt

\bio{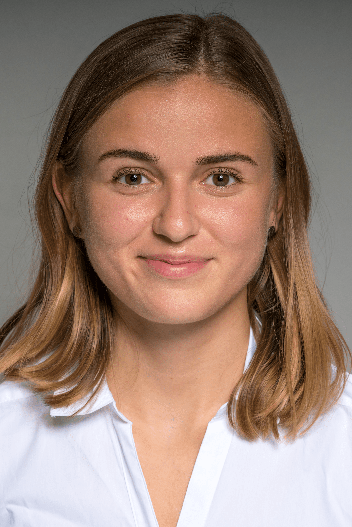}
Andrine Elsetrønning is a student at the Department of Engineering Cybernetics at the Norwegian University of Science and Technology, intending to finish her master degree during the spring of 2021. The research presented in the proposed paper, is based on the research conducted as a part of her pre-thesis specialization project.  
\endbio
\newpage
\bio{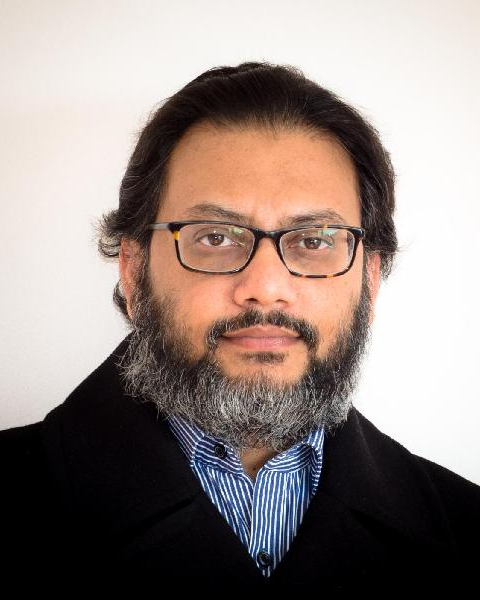}
Adil Rasheed is the professor of Big Data Cybernetics in the Department of Engineering Cybernetics at the Norwegian University of Science and Technology where he is working to develop novel hybrid methods at the intersection of big data, physics driven modelling and data driven modelling in the context of real time automation and control. He also holds a part time senior scientist position in the Department of Mathematics and Cybernetics at SINTEF Digital where he led the Computational Sciences and Engineering group between 2012-2018. He holds a PhD in Multiscale Modeling of Urban Climate from the Swiss Federal Institute of Technology Lausanne. Prior to that he received his bachelors in Mechanical Engineering and a masters in Thermal and Fluids Engineering from the Indian Institute of Technology Bombay.
\endbio

\bio{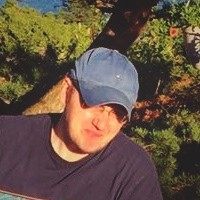}
Jon Bekker is a medical doctor and entrepreneur. He is interested in improving healthcare with cross-discipline knowledge and collaboration. He has founded dedeX a medtech company with the aim of using artificial intelligence to improve our interpretation of the medical history and physical examination.
\endbio
\bio{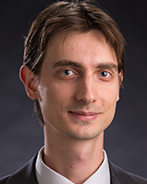}
Omer San received his bachelors in aeronautical engineering from Istanbul Technical University in 2005, his masters in aerospace engineering from Old Dominion University in 2007, and his Ph.D. in engineering mechanics from Virginia Tech in 2012. He worked as a postdoc at Virginia Tech from 2012-'14, and then from 2014-'15 at the University of Notre Dame, Indiana. He has been an assistant professor of mechanical and aerospace engineering at Oklahoma State University, Stillwater, OK, USA, since 2015. He is a recipient of U.S. Department of Energy 2018 Early Career Research Program Award in Applied Mathematics. His field of study is centered upon the development, analysis and application of advanced computational methods in science and engineering with a particular emphasis on fluid dynamics across a variety of spatial and temporal scales. 
\endbio

\end{document}